\documentclass[twoside]{ilcws08}
\usepackage[latin1]{inputenc}
\pdfoutput=1
\usepackage{epsfig}
\usepackage{graphicx}
\usepackage{color}
\usepackage{wrapfig,rotating}
\usepackage{amssymb,amsmath,array}

\begin{document}
\title{
%%%%   Paper title goes here  %%%%%%%%%%%%%%
A Large TPC Prototype with MPGD Readout: Status and Plans} %% 
%***********************************************************************
% AUTHORS INFORMATION AREA
%***********************************************************************
\author{Ties Behnke$^1$, Klaus Dehmelt$^1$, Takeshi Matsuda$^{1,2}$ and Peter Schade$^1$
% Optional short acknowledgment: remove next line if non-needed
%\thanks{This is an optional funding source acknowledgment.}
% DO NOT MODIFY THE FOLLOWING '\vspace' ARGUMENT
\vspace{.3cm}\\
% Addresses and institutions (remove "1- " in case of a single institution)
1- DESY-FLC, Notkestrasse 85, 22603 Hamburg, Germany \\
%% Remove the next three lines in case of a single institution
\vspace{.1cm}\\
2- High Energy Accelerator Research Organization, KEK, 1-1 Oho, Tsukuba, Ibaraki 305-0801, Japan\\
}
%%***********************************************************************
% END OF AUTHORS INFORMATION AREA
%***********************************************************************

\maketitle

\begin{abstract}
The use of a TPC in future collider experiments needs significant research and development. Within the EUDET program an infrastructure has been designed and built to allow the efficient testing of TPC prototypes under realistic conditions. The infrastructure consists of a test beam facility, located at DESY Hamburg, and a multi-purpose TPC field cage. 
\end{abstract}

\section{Introduction}
Time projection chambers have been successfully used in many experiments at colliding beam facilities. The latest example is the ALICE TPC, which has very recently been commissioned, and which is scheduled to start recording 
collisions as soon as the LHC commences operation. 
The next generation of TPCs is now planned for future collider projects, 
for example the international linear collider. These future chambers need to 
deliver excellent precision, need to operate stable and reliable, and 
will need to operate in a high luminocity collider where many bunch collisions take place during one readout cycle of a TPC. An important criterion in addition is 
the amount of material presented by the detector, which should be minimal. 

A traditional TPC based on a wire chamber readout does to 
meet all requirements. Alternatives solutions have been studied over the past 
few years, which are based on readout systems using micro pattern gas detectors, e.g. Gas Electron Multipliers (GEM) or Micro Mesh Gas Amplifiers (MicroMegas). 
Both promise to deliver a potentially better spatial resolution, due to the 
smaller intrinsic size of the gas amplification structures, they offer 
stable and simple operation, and they promise in addition to operate 
even in a multi-bunch crossing scenario like the ILC since they naturally
suppress the amount of ions back-drifting into the TPC volume.

\section{The EUDET TPC facility}
The LC-TPC collaboration~\cite{LCTPC} has been formed to develop the case for a TPC 
at the linear collider. In the past the feasibility of a MPGD based TPC 
has been successfully demonstrated, both for GEM and MicroMegas readout, 
using small scale prototypes. The next step now is a larger system test 
of a TPC with beam, with a significant number of readout channels, to 
prove that the resolution and overall performance requirements of a TPC at the ILC can be met. 

The EUDET project~\cite{EUDET} is developing a test infrastructure, which is well suited for 
the needs of the LC-TPC collaboration. It consists of a beam line, located 
at the electron test beam facility at DESY Hamburg, equipped with a 
1~T superconducting magnet, and needed auxiliary equipment. It also 
includes a multi-purpose field cage built to be used with 
different readout modules for a TPC. The field cage is designed 
such that it fits into the magnet, and can optimally exploit the 
EUDET facility. 

The LC-TPC collaboration is the first major user of this facility. It is 
equipping the field cage with two types of readout systems (GEM and 
MicroMegas), and will test several different readout electronics systems 
with the same chamber. The collaboration is also contributing significantly 
to the commissioning and further development of the EUDET infrastructure. 

\section{Status of the superconducting Magnet}
The EUDET facility is equipped with a superconducting magnet, which can provide 
a solenoidal field of around 1.0~T. The magnet is on lease from KEK, where it 
has been developed for a balloon based experiment. It has a number of 
special features, which make it particularly well suited for the 
EUDET facility. Its construction has been optimized for low mass. The 
amount of material in the walls of the magnet is less than $10\%$ of 
a radiation length per side, which makes it possible that an electron beam 
can penetrate the walls of the magnet and still be used for a test 
inside the magnet. The magnet is also designed so that it can be 
moved while cold, so that the magnet as a whole can be placed on 
a movable stage in the beam, and the complete active area of the 
chamber inside the magnet can be scanned with beam. Due to the 
requirements of low mass, the magnet is not equipped with an 
iron return yoke. This has a significant impact on the 
field quality inside the magnet. 

\begin{wrapfigure}{r}{0.5\columnwidth}
\centerline{\includegraphics[clip,width=0.45\columnwidth,viewport={0 0 11.03in 8.26in}]{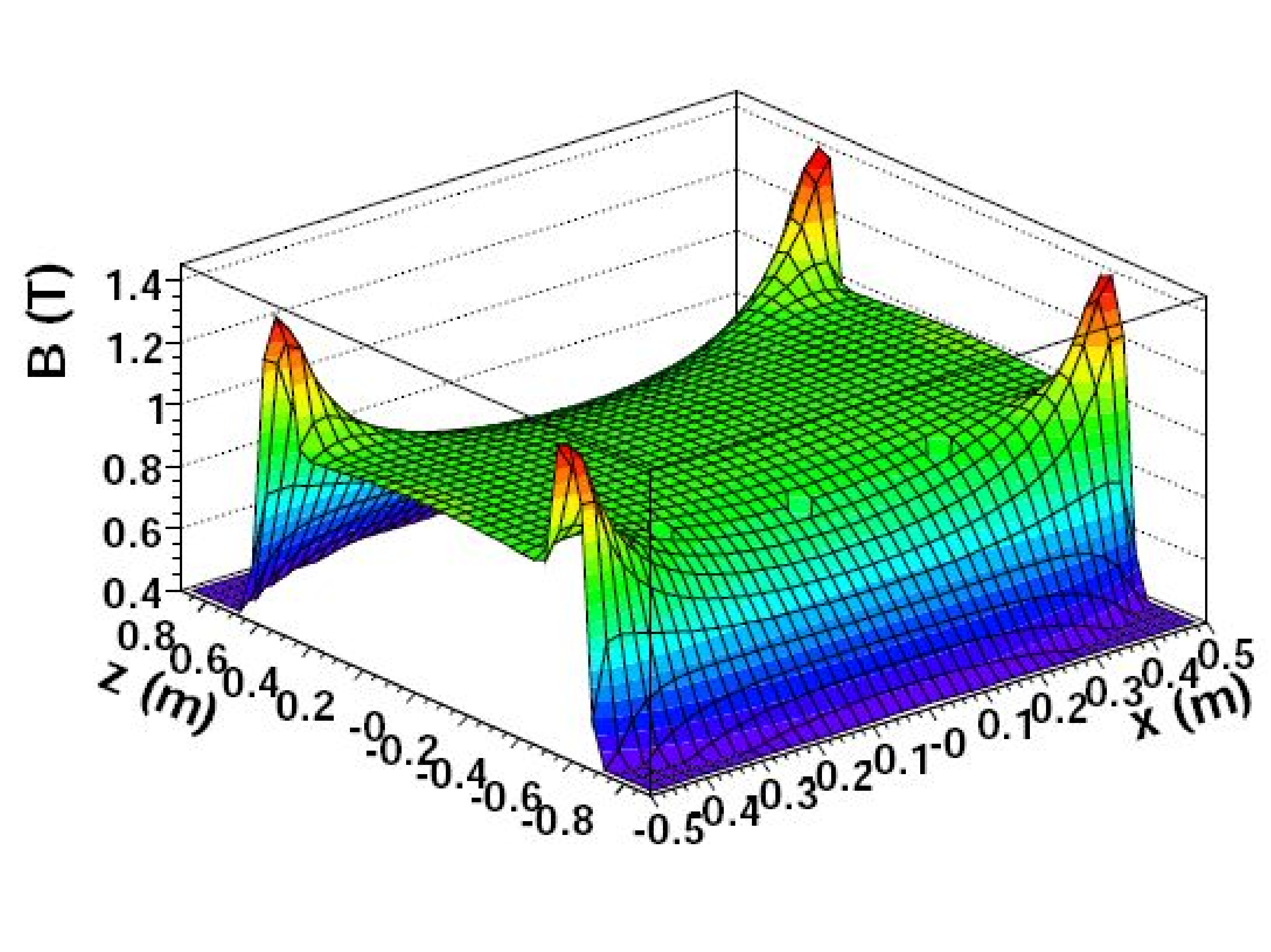}}
\caption{Breakdown of the different materials used in the construction of the TPC field cage.}\label{fig:fieldmap}
\end{wrapfigure}

The magnet was delivered to DESY in 2007, and installed and 
commissioned in 2008. First experiment using the magnet were 
conducted at the end of 2008. 

An important step was the precise measurement of the magnetic 
field map of the magnet. Using a system of hall probes mounted 
on a precision arm, several thousand points were recorded inside the 
magnet volume. The resulting field map was parametrized 
based on a model of the expected field shape~\cite{Grefe}. In Figure~\ref{fig:fieldmap} 
the result of the parametrization is shown. The variations of the 
field are as large as 0.2~T. Point by point, the largest deviations between 
the model and measurements are as much as 10~G, or $10^{-3}$. This is 
somewhat worse than the goal of a few times $10^{-4}$, but still 
good enough for most of the planned measurements.

At the moment the magnet is installed in a fixed position in the 
beam line at DESY. In the first half of 2009 the area will be 
equipped with a remote controllable table, which can move the 
magnet vertically, horizontally through the beam, and also 
rotate the magnet relative to the beam by up to $40$degree. 

\section{Field Cage}
\subsection{Overall program goals}
Testing different readout mechanisms of a TPC under comparable conditions 
require that a multi-purpose TPC is built, which can equally well operate
with different readout systems. The goal of the LC-TPC collaboration 
together with the EUDET program has been to develop such a multi
purpose infrastructure. A central part of this is a field cage, 
which fits inside the magnet described in the previous section, 
and which can be equipped with a wide range of different 
readout modules.

\subsection{Field Cage Construction}
The field cage is a light-weight cylindrical structure. Its main parameters 
are summarized in table~\ref{tab:fieldcage}. 

\begin{wraptable}{l}{0.5\columnwidth}
\centerline{\begin{tabular}{|l|c|}
\hline
Parameter  & value \\\hline
Outer Diameter (cm)& 77 \\
Inner active Diameter (cm)& 72\\
Length (cm)        & 61\\
Drift Length (cm)  & 58 \\
Wall thickness (\% X$_0$) & 1.3\\
Cathode & Al disk \\
\hline
\end{tabular}}
\caption{Basic parameters of the TPC prototype.}
\label{tab:fieldcage}
\end{wraptable}

\begin{wrapfigure}{r}{0.5\columnwidth}
\centerline{\includegraphics[clip,width=0.45\columnwidth,viewport={5 5 411 257}]{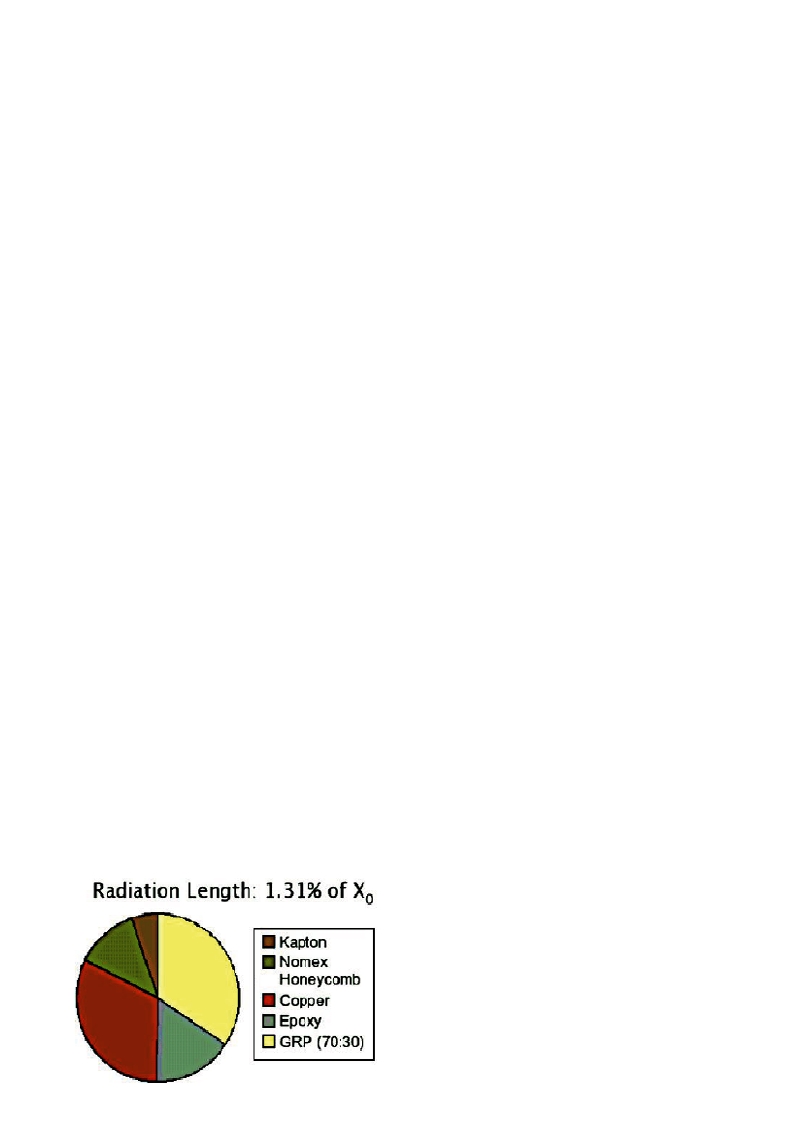}}
\caption{Breakdown of the different materials used in the construction of the TPC field cage.}\label{fig:material}
\end{wrapfigure}

The walls are a light 
weight composite structure, build around a core of Nomex with thin 
skins of glas-fibre reinforced epoxy on either side. Electrically 
Kapton sheets provide both insulation and - on the inside - 
a pattern of field shaping strips. Overall one wall of the 
TPC corresponds to $1.3\%$ of a radiation length (see Figure~\ref{fig:material}). 
The largest 
contribution comes from the epoxy/ glas fibre system and from the copper used for 
the field shaping strips and a grounding layer on the outside~\cite{Schade}. 

The two ends of the cylinder are terminated with flanges, built from a 
high density foam. The end plate and the cathode plate are attached to these
flanges with a ring of M6 stainless steel screws. An O-ring 
ensures the necessary gas tightness. The cathode is a disk of aluminum copper coated on the inside. It is supported at three points from 
a lid which is used to close off the field cage. The three point support is 
designed in a way that the cathode can be aligned parallel to the 
field strips, to ensure an optimal field quality inside the field cage. 

The building of the field cage was done by an outside company over a 
purpose built mandrel. A careful survey of the field cage after the 
construction showed that the dimensions of the system were 
within $0.1$~mm of the specified parameters. The two end flanges 
are parallel to each other to better than $0.2$~mrad. The flanges are 
however at a slight angle relative to the field cage of $1$~ mrad. This 
corresponds to a relative displacement of the axes of the two end plates to 
each other of $0.54$~mm, which is outside the tolerance of $0.1$~mm. The 
impact of this on the ultimate precision with which the electric 
field in the chamber is know, is under evaluation. 

\begin{wrapfigure}{r}{0.5\columnwidth}
\centerline{\includegraphics[clip,width=0.45\columnwidth,viewport={15 0 2400 1600}]{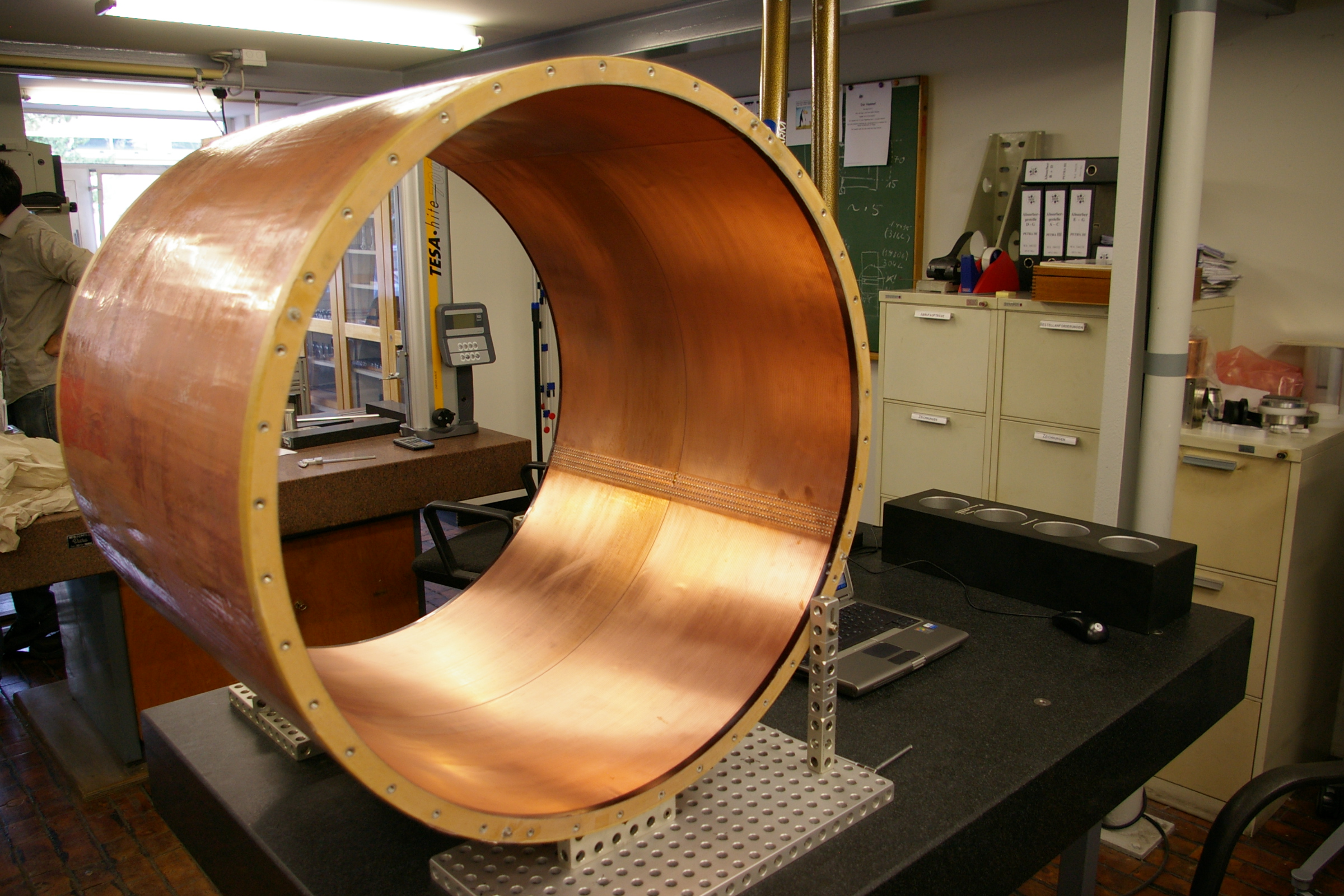}}
\caption{Picture of the finished TPC fieldcage, without endplates attached.}\label{fieldcage}
\end{wrapfigure}

A picture of the finished 
field cage is shown in figure~\ref{fieldcage}.

\subsection{TPC End Plate}
The readout modules of the TPC are help by an aluminum end plate, 
which is described in more detail elsewhere. In the first iteration the 
end plate has been built from aluminum, with openings for seven readout modules. The 
readout modules are arranged on a circle which corresponds to the 
expected circle in the final TPC at the ILC. Frames machined from 
Aluminum are pulled against this end plate from the inside of the 
TPC, so that they cover the complete area, without any dead area. 
The readout modules are then built up on these frames. A system of 
dowel pins ensures that the frames are precisely positioned 
relative to the end plate. 

A major challenge of the TPC developments for the linear collider is the 
minimization of material in the end plate. The Aluminum model clearly 
is not optimized in this direction, but serves as a first and 
stable platform to develop and test readout modules. It is 
planned to replace the end plate by a lighter one in a second step. 
The design of this lighter end plate is not yet finished. Possibly it 
will be based on Carbon-fibre technology, or another advanced 
composite material system. A reduction of the material in the 
endplate of at least a factor of two is envisioned. 

\subsection{Commissioning of the TPC}
After delivery of all parts the TPC was commissioned during the fall of 2008. 
After initial assembly and gas tightness and high voltage tests, the 
chamber was equipped with a single sensitive module, based on the 
MicroMegas technology. More details about the readout module may be 
found in \cite{LP-COLAS}. No major problems were encountered during the 
commissioning phase. The field cage stood the high voltage without problems. 
The gas tightness of the system was achieved with minimal problems. 

\section{Conclusion}
A light weight field cage for a TPC has been designed and built. It is equipped 
with a readout system which can support up to seven readout modules. 
A total material equivalent of $1.3\%$ of a radiation length has been 
achieved for one wall of the field cage. Excellent mechanical 
properties were obtained. Equipped with an aluminum end plate the 
chamber was commissioned in the fall of 2008 without major problems. 
It will be operated at the DESY electron test beam for the next 
years equipped with different readout modules based on a range of 
different technologies. 

\section{Acknowledgments}
The construction and building of the TPC infrastructure is the 
result of a collaboration of many different people, whose support and 
help in writing this report is acknowledged.  
%\begin{wrapfigure}{r}{0.5\columnwidth}
%\centerline{\includegraphics[width=0.45\columnwidth]{ilcws08.eps}}
%\caption{ILC Logo}\label{Fig:MV}
%\end{wrapfigure}


\begin{thebibliography}{99}
\bibitem{LCTPC} see {\it http://www.lctpc.org}
\bibitem{EUDET} see {\it http://www.eudet.org}
\bibitem{Grefe}{C.~Grefe, ``Magnetic field map for a large TPC prototype,'', DESY-THESIS-2008-052}
\bibitem{Schade}{P.~Schade {\it et.al.}, "Status and Plans of the Large TPC Prototype for the ILC", EUDET-Memo-2007-37}
\bibitem{LP-COLAS}{P. Colas, in these proceedings}
\end{thebibliography}
\end{document}